\pdfoutput=1
\documentclass{revtex4}
\usepackage{latexsym}
\usepackage{amsfonts}
\usepackage[latin1]{inputenc}
\usepackage[caption=false]{subfig}
\usepackage{graphicx}
\usepackage{mathrsfs}
\usepackage{amssymb}
\usepackage{amsmath}
\usepackage{color}
\usepackage{epsfig}
\usepackage{fancyhdr}
\usepackage{color}

\begin{document}
\begin{flushright}
PSI-PR-14-08
\end{flushright}

\title{Composite Higgs searches at the LHC and beyond}

%

\author{D. Barducci, A. Belyaev, M.S.Brown, S. Moretti}
\affiliation{School of Physics and Astronomy, University of Southampton, Highfield, SO17 1BJ, UK.}
\email{d.barducci,a.belyaev,m.s.brown@soton.ac.uk,s.moretti@soton.ac.uk}
\author{S. De Curtis}
\affiliation{INFN, Sezione di Firenze,
Via G. Sansone 1, 50019 Sesto Fiorentino, Italy.}
\email{decurtis@fi.infn.it}
\author{G.M. Pruna}
\affiliation{Paul Scherrer Institute, CH-5232 Villigen PSI, Switzerland.}
\email{giovanni-marco.pruna@psi.ch}

\begin{abstract}
General Composite Higgs models provide an elegant solution to the hierarchy problem present in the Standard Model
and give an alternative pattern leading to the mechanism of electroweak (EW) symmetry breaking.
We present an analysis of a realistic realization of this general idea, namely the 4DCHM, analysing
the Higgs production and decay modes, fitting them to the latest LHC showing the compatibility with the results of the CERN machine.
We then present the prospects of a future electron positron collider
of testing this model against the expected experimental accuracies in the various Higgs decay channels accessible herein. 
\end{abstract}

\maketitle

\thispagestyle{fancy}


\section{Introduction}

After the discovery of a Higgs like state at the mass of 125 GeV made at CERN from the ATLAS \cite{Aad:2012tfa} and CMS \cite{Chatrchyan:2012ufa} collaborations, one of 
the primary questions that the physics community ought to answer is whether this particle is consistent with the one
predicted by the Standard Model (SM).
From an experimental point of view the properties of this particle show both agreement
and tensions with the predictions of the SM, although the errors on these measurements do not allow yet to draw a final conclusion on its properties.
Conversely, from a theoretical point of view there are many motivations to think that the SM is not the ultimate
and complete theory of Nature, among which the naturalness argument plays a predominant role.
The instability of the Higgs mass with respect to radiative corrections requires in fact an
incredible high level of fine tuning in the precision of their cancellation  in the SM in order to have an Higgs mass at the EW scale.
Beside the supersymmetric solution to this problem, another possibility is to postulate the Higgs boson as a composite 
state arising as a bound state from a strongly interacting sector at the TeV scale \cite{Kaplan:1983fs}.
Being composite the Higgs will be insensitive to radiative corrections above the composite scale and its lightness
with respect to other resonances of the strong sector can be taken into account by postulating the Higgs as a pseudo Nambu Goldstone
boson (pNGB), similarly to what happens for the pions in QCD.
Among the various composite Higgs models present in literature, we will show in this proceeding the compatibility of a recent
proposed framework, the 4D Composite Higgs Model (4DCHM) of \cite{DeCurtis:2011yx} based on a $SO(5)/SO(4)$ breaking pattern, with respect to the 7 and 8 TeV LHC data and the capabilities of a future proposed $e^+e^-$ collider in unravelling the composite nature
of the 125 GeV scalar boson. See Refs. \cite{Barducci:2013wjc,Barducci:2013ioa} for further details.

\section{LHC analysis}

\begin{figure}[!ht]
\begin{picture}(17,160)
\put(-194.3,5.7){\epsfig{file=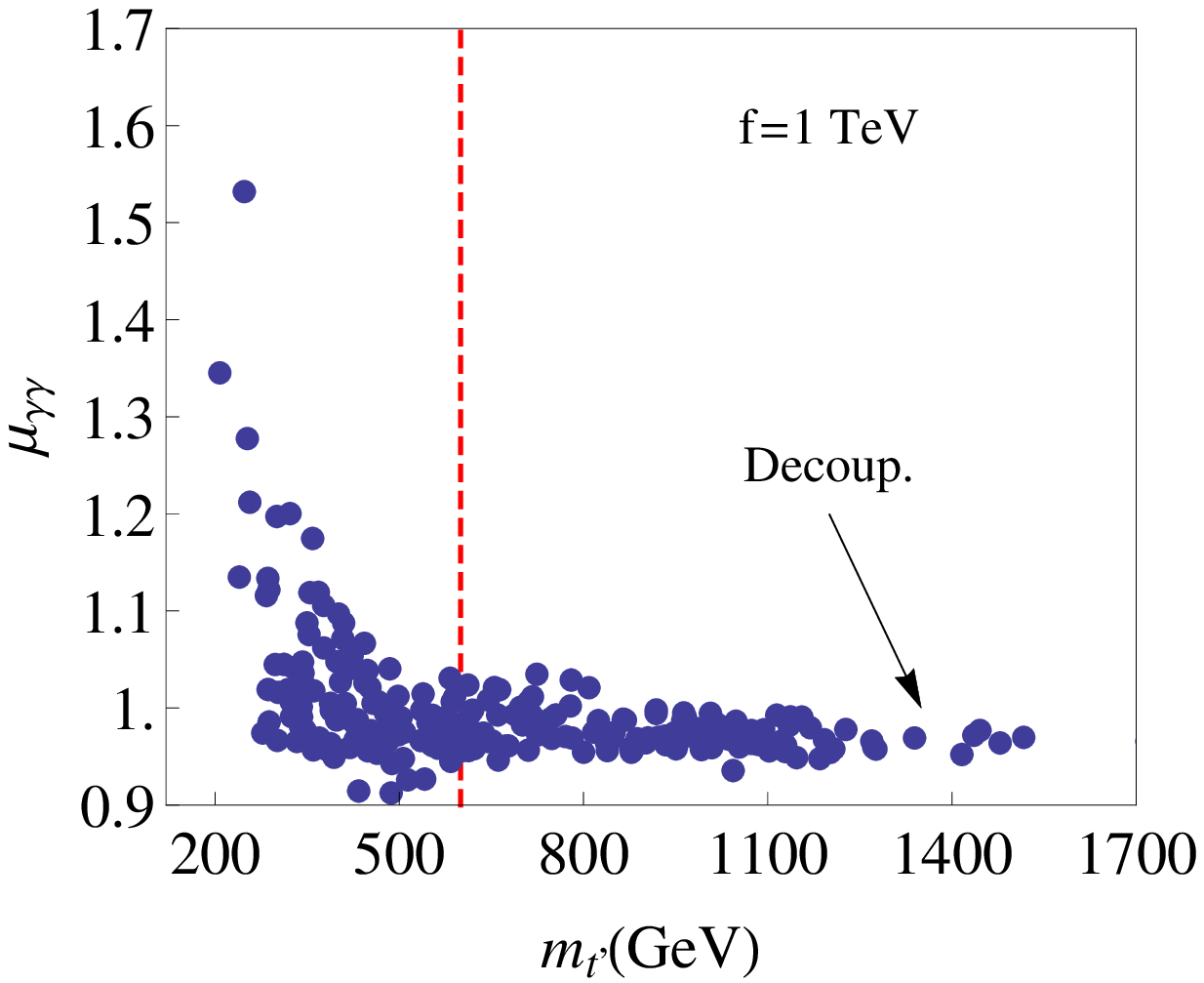, width=.40\textwidth}{(a)}}
\put(34.3,15.7){\epsfig{file=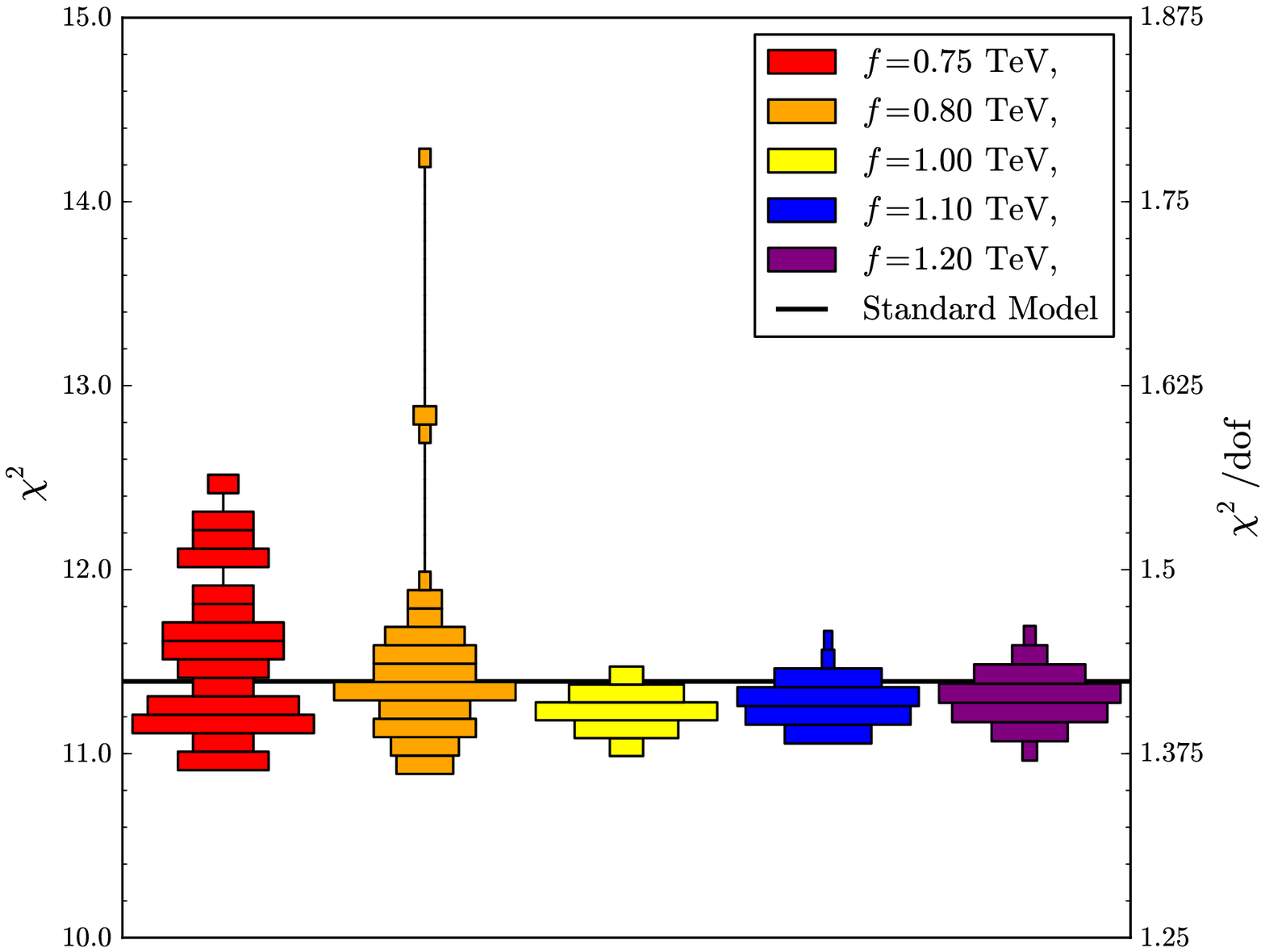, width=.40\textwidth}{(b)}}
\end{picture}
\caption{$\gamma\gamma$ signal strenght in function of the mass of the lightest extra quark with charge 2/3 (a) and $\chi^2$ fit for various model scales (b) with the 7 and 8 TeV LHC data.}
\label{fig:lhc}
\end{figure}

The couplings of the 4DCHM Higgs to SM model gauge bosons and fermions are modified with respect to the SM ones through factors that depend on the model scale $f$, explicitly $g_{ffH}/g^{SM}_{ffH}=(1-2\xi)/\sqrt{1-\xi}$ and $g_{VVH}/g^{SM}_{VVH}=\sqrt{1-\xi}$, with $\xi=v^2/f^2$ and where $v$ is the Higgs vacuum expectation value. Beside these modifications due to the pNGB nature, other source of modifications are due to mass mixing between SM fermions and gauge bosons with the extra ones present in the 4DCHM and, in case of loop induced processes, of extra particles that can run inside the loops. Taking into account all these aspects we plot in Fig.~\ref{fig:lhc} (a) the prediction for the $\gamma\gamma$ signal strength, $\mu_{\gamma\gamma}$ (that is the ratio of the 4DCHM event rate with respect to the SM expectation), in function of the mass
of the lightest extra top present in the model, for a model scale of 1 TeV and showing, with a vertical dashed red line, an approximate exclusion limit on the masses of the extra fermions obtained by recasting the available results for direct searches of top partners. The arrow indicates the expected 4DCHM signal strenght in case of the decoupling of all the extra particle content of the model. Comparing then our predictions for the signal strenght in the $b\bar b$, $WW$, $ZZ$ and $\gamma\gamma$ channels we have performed a $\chi^2$ fit, Fig.~\ref{fig:lhc} (b), for various model scale choices, where we can see the compatibility of the 4DCHM with the latest LHC Higgs data, also with respect to the SM, represented by an horizontal black line.

\section{Future $e^+e^-$ collider analysis}

\begin{figure}[!ht]
\begin{picture}(17,160)
\put(-74.3,5.7){\epsfig{file=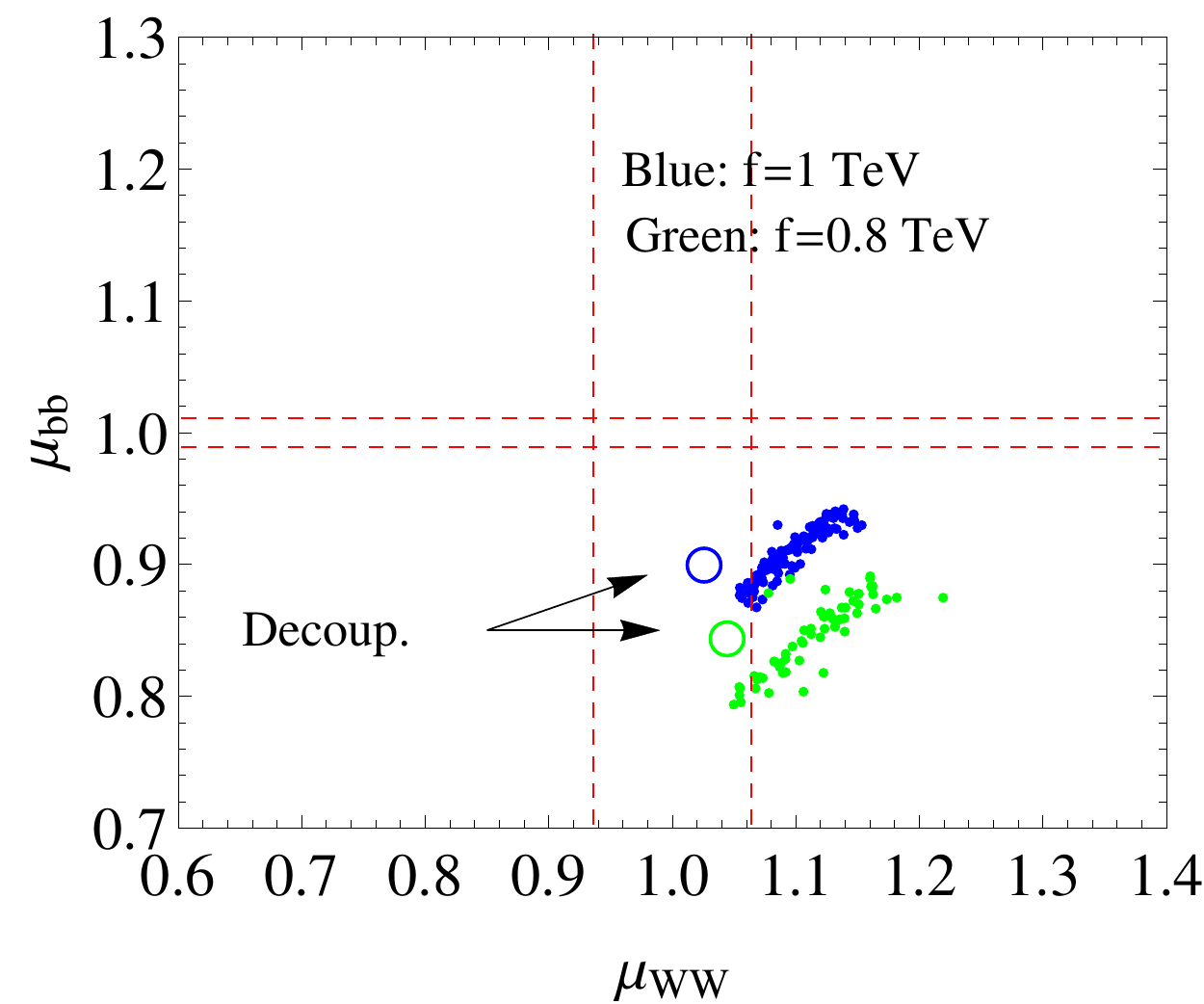, width=.40\textwidth}}
\end{picture}
\caption{$\mu_{bb}$ and $\mu_{WW}$ signal strengths for the Higgs-strahlung production process at the ILC with $\sqrt{s}$=250 GeV for $f=1$ TeV (blue) and $f=0.8$ TeV (green). Red dashed lines represent the expected experimental accuracies in measuring these observables according to \cite{Baer:2013cma} while the circles represent the 4DCHM predictions in case of the decoupling of all the extra particle content.}
\label{fig:lc}
\end{figure}

We then tested our framework against a future $e^+e^-$ collider for which we have chosen as a benchmark the proposed International Linear Collider (ILC) \cite{Baer:2013cma}.
We have analysed the 4DCHM signal strengths and compared them with the predicted accuracies for this machines in measuring these observables.
In case of Higgs-strahlung production at $\sqrt{s}=250$ GeV
we plot the values of $\mu_{bb}$ and $\mu_{WW}$ for two different model scales, $f=1$ TeV and $f$=0.8 TeV, in Fig.~\ref{fig:lc}, where we also show with dashed red lines the expected experimental accuracies for this observables with 250 fb$^{-1}$ of integrated luminosity and with the circles the values of the signal strengths in the limit of the decoupling
of all the extra particle content of the 4DCHM.
From the plot we observe that the ILC has the potential, already at $\sqrt{s}=250$ GeV, of disentangling the 4DCHM with respect to the SM hypothesis and also the importance of keeping the full particle spectrum of the model, which would indeed render the deviations from the SM manifest, in contrast to the decoupling limit, which would then be clearly inappropriate to adopt in this case.

\section{Conclusions}

In conclusion in this proceeding we have shown that the 4DCHM shows compatibility with the LHC data pointing to 
the discovery of a Higgs boson at 125 GeV and that a future $e^+e^-$ collider will be able to test with higher precision
the properties of this state so as to understand its nature.
\acknowledgments
DB, AB and SM are financed in part through the NExT Institute. The work of GMP has been supported by the European Community's Seventh
Framework Programme (FP7/2007-2013) under grant agreement n.~290605
(COFUND: PSI-FELLOW). 

\end{document}